\documentclass{ws-ijmpcs}
\begin{document}

\markboth{Zhongbao Yin for the ALICE Collaboration}
{Strange and multi-strange particle production in Pb-Pb collisions at $\sqrt{s_{NN}}$ = 2.76 TeV with ALICE}

\catchline{}{}{}{}{}
%
%%%%%%%%%%%%%%%%%%%%%%%%%%%%%%%%%%%%%%%%%%%%%%%%%%%%%%%%%%%%%%%%%%%%

\title{STRANGE AND MULTI-STRANGE PARTICLE PRODUCTION IN Pb-Pb COLLISIONS AT $\sqrt{s_{\rm{NN}}}$ = 2.76 TeV WITH ALICE}

\author{ZHONGBAO YIN, FOR THE ALICE COLLABORATION}

\address{Key Laboratory of Quark and Lepton Physics (MOE) and Institute of Particle Physics, 
Central China Normal University, Wuhan 430079, China\\
zbyin@mail.ccnu.edu.cn}

\maketitle

\begin{history}
\received{Day Month Year}
\revised{Day Month Year}
\end{history}

\begin{abstract}
We present ALICE results on strange and multi-strange hadron production  
as a function of centrality in Pb-Pb collisions 
at $\sqrt{s_{\rm{NN}}}$ = 2.76 TeV at the LHC. 
Their transverse momentum spectra, 
yields and particle ratios are compared to the corresponding measurements 
in pp collisions to address strangeness enhancement and 
high $p_{\rm T}$ particle suppression. The results are also compared 
to measurements at RHIC and SPS energies.

\end{abstract}

\ccode{PACS numbers: 25.75.-q, 14.20.Jn, 14.40.-n, 21.65.Qr}

\section{Introduction}	
It is generally assumed that collisions of relativistic heavy-ions 
lead to the formation of a deconfined high temperature and density state of nuclear matter, 
the quark-gluon plasma. The new phase of matter exists for a short 
time after the collisions and ultimately hadronizes into final-state 
particles. Considerable efforts have been put into identifying 
observables that are sensitive to the properties of 
the early deconfined stage of the collisions. Strangeness enhancement, 
one of the first proposed signatures of the deconfined phase\cite{rafe82},   
has been commonly considered to be an important probe of 
the strongly interacting matter created in heavy-ion collisions.

By exploiting the tracking and particle identification capabilities of ALICE,
strange ($\Lambda$, K$^0_{\rm S}$) and multi-strange ($\Xi^{-}$, $\Omega^{-}$)
particles can be detected via their weak decay topologies and 
measured over a wide range of transverse momenta. In this report we 
present the latest ALICE results\footnote{Some of the results presented 
at the workshop have, in the meantime, been finalized and these results 
are used in these proceedings.} 
on strange ($\Lambda$, K$^0_{\rm S}$) and 
multi-strange ($\Xi^{-}$, $\Omega^{-}$) production in Pb-Pb collisions at $\sqrt{s_{\rm{NN}}} = 2.76$ TeV. 
In the next section, 
we present transverse momentum ($p_{\rm T}$) spectra measured at 
mid-rapidity ($|y| < 0.5$) as function of centrality. 
We discuss in the third section the evolution of $\Lambda/\rm{K}^0_{\rm S}$ 
ratio as a function of $p_{\rm T}$, in comparison with the corresponding 
results in pp collisions at the LHC and in $\sqrt{s_{\rm {NN}}} = 200$ 
GeV Au-Au collisions at RHIC. The excitation function of 
the strangeness enhancement from SPS to LHC energies 
is discussed in the fourth section. In the fifth section, 
the nuclear modification factors for multi-strange particles are 
compared with other identified particles to provide insight into 
particle production and energy loss mechanisms at play.
 
\section{Strange and Multi-strange $p_{\rm T}$ Spectra in Pb-Pb Collisions 
at $\sqrt{s_{\rm {NN}}} = 2.76$ TeV}

The ALICE experiment\cite{aamo08}, dedicated to study heavy-ion physics 
at the LHC, has excellent particle identification capability. 
Strange and multi-strange particles were 
reconstructed via their weak decay
channels: $\rm{K}^{0}_{\rm S} \rightarrow \pi^{+}\pi^{-}$, 
$\Lambda\rightarrow \rm{p}+\pi^{-}$
($\bar{\Lambda}\rightarrow \bar{\rm{p}} + \pi^{+}$), 
$\Xi^{-} \rightarrow \Lambda +\pi^{-}$
($\bar{\Xi}^{+} \rightarrow \bar{\Lambda} +\pi^{+}$)
and $\Omega^{-} \rightarrow \Lambda +\rm{K}^{-}$
($\bar{\Omega}^{+} \rightarrow \bar{\Lambda} +\rm{K}^{+}$).
The combinatorial background contribution to the
invariant mass distributions of the candidates for all species 
was reduced by applying cuts selecting specific decay topologies 
and using information on the specific energy loss in TPC 
of their decay daughters. 

\begin{figure}[!htb]
\begin{minipage}[b]{0.49\linewidth}
\centerline{\epsfig{file=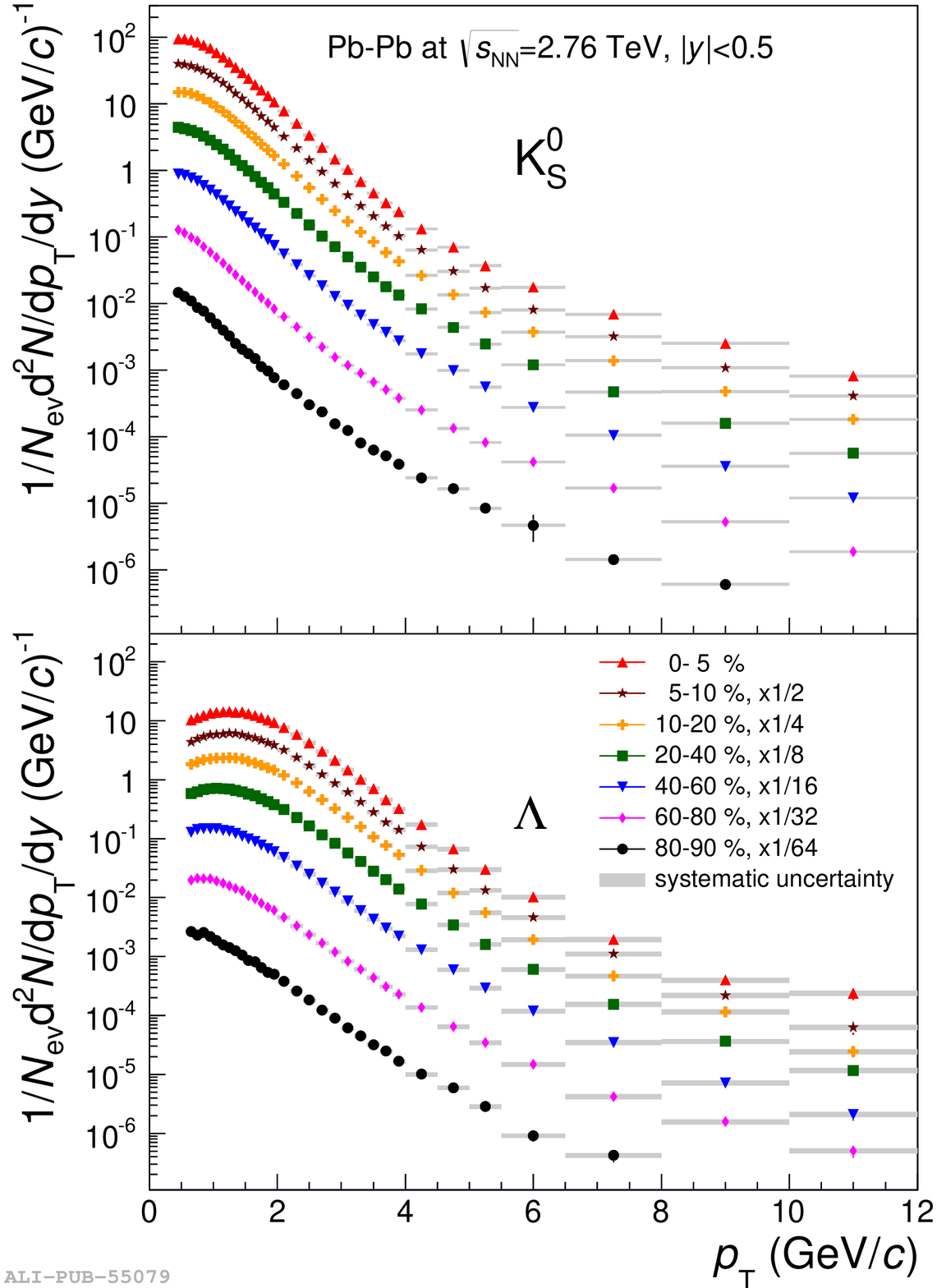, width=6.cm}}
\end{minipage}
\hspace{0.cm}
\begin{minipage}[b]{0.49\linewidth}
\centerline{\epsfig{file=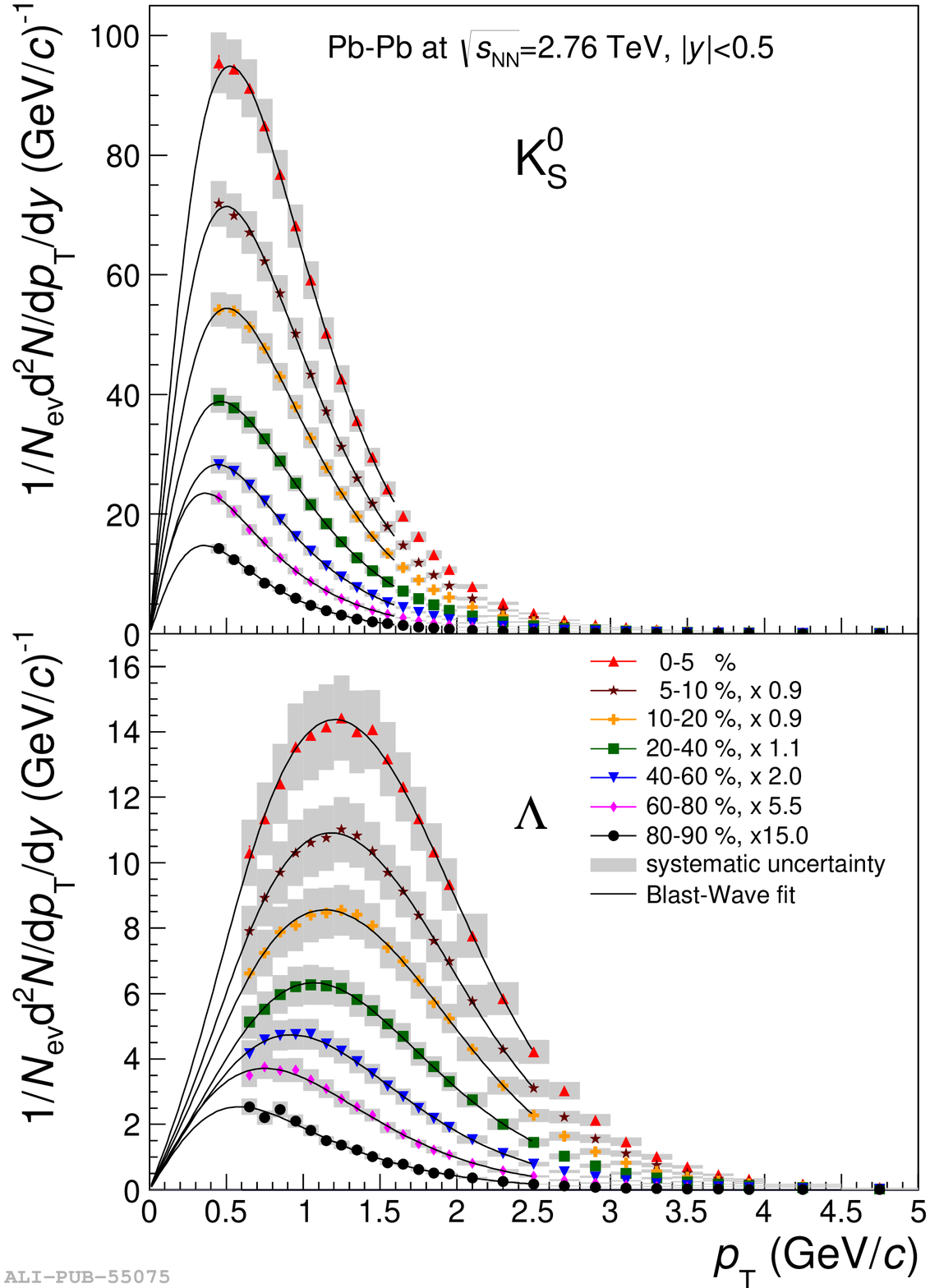, width=6.0cm}}
\end{minipage}
\vspace*{8pt}
\caption{Transverse momentum spectra for $\rm{K}^{0}_{\rm S}$ (top) 
and $\Lambda$ (bottom) at mid-rapidity for different centrality intervals 
of Pb-Pb collisions shown in logarithmic (left) and linear (right) scale. 
\label{fig1}}
\end{figure}

\begin{figure}[!hbt]
\begin{minipage}[b]{0.49\linewidth}
\centerline{\epsfig{file=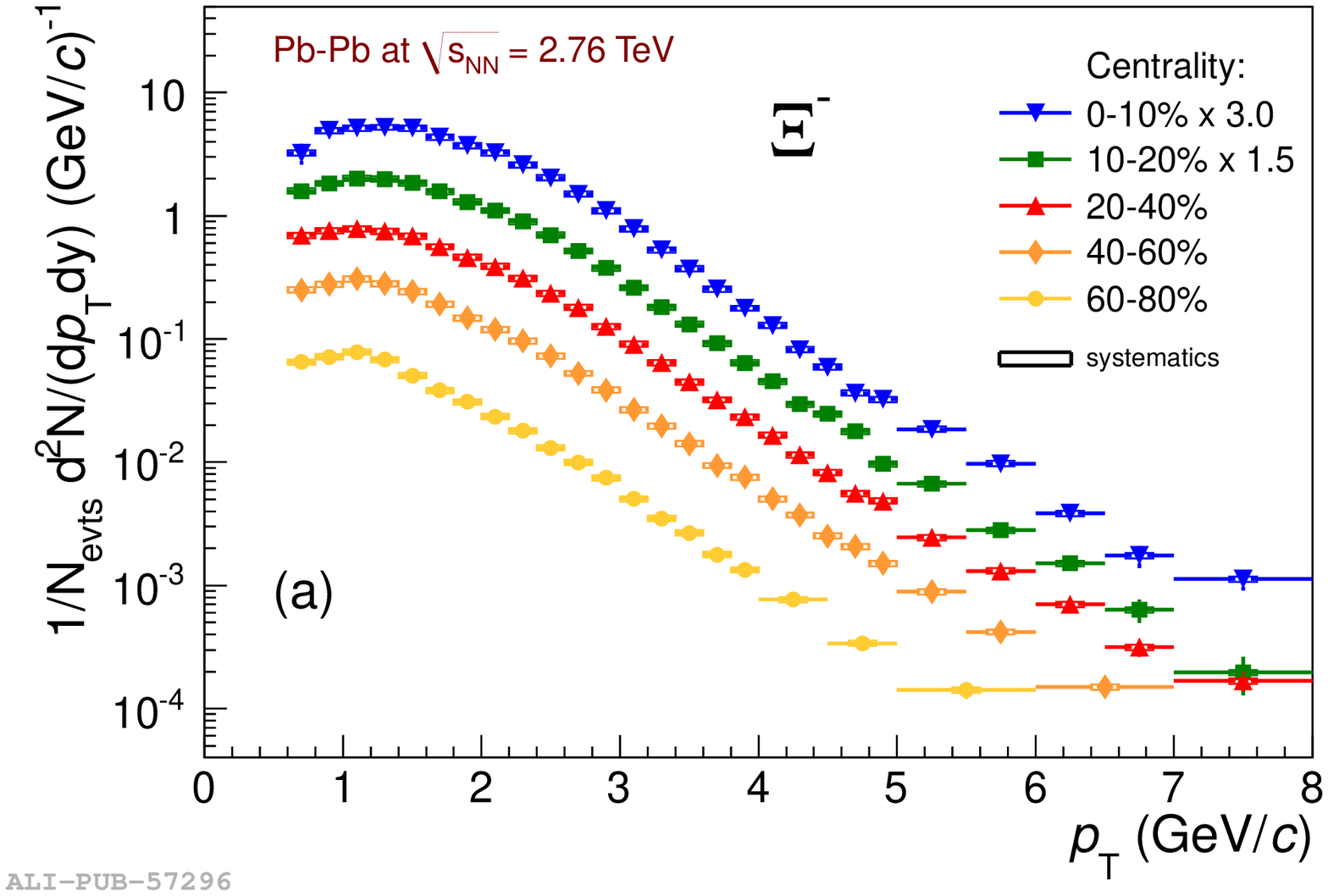, width=6.cm}}
\end{minipage}
\hspace{0.cm}
\begin{minipage}[b]{0.49\linewidth}
\centerline{\epsfig{file=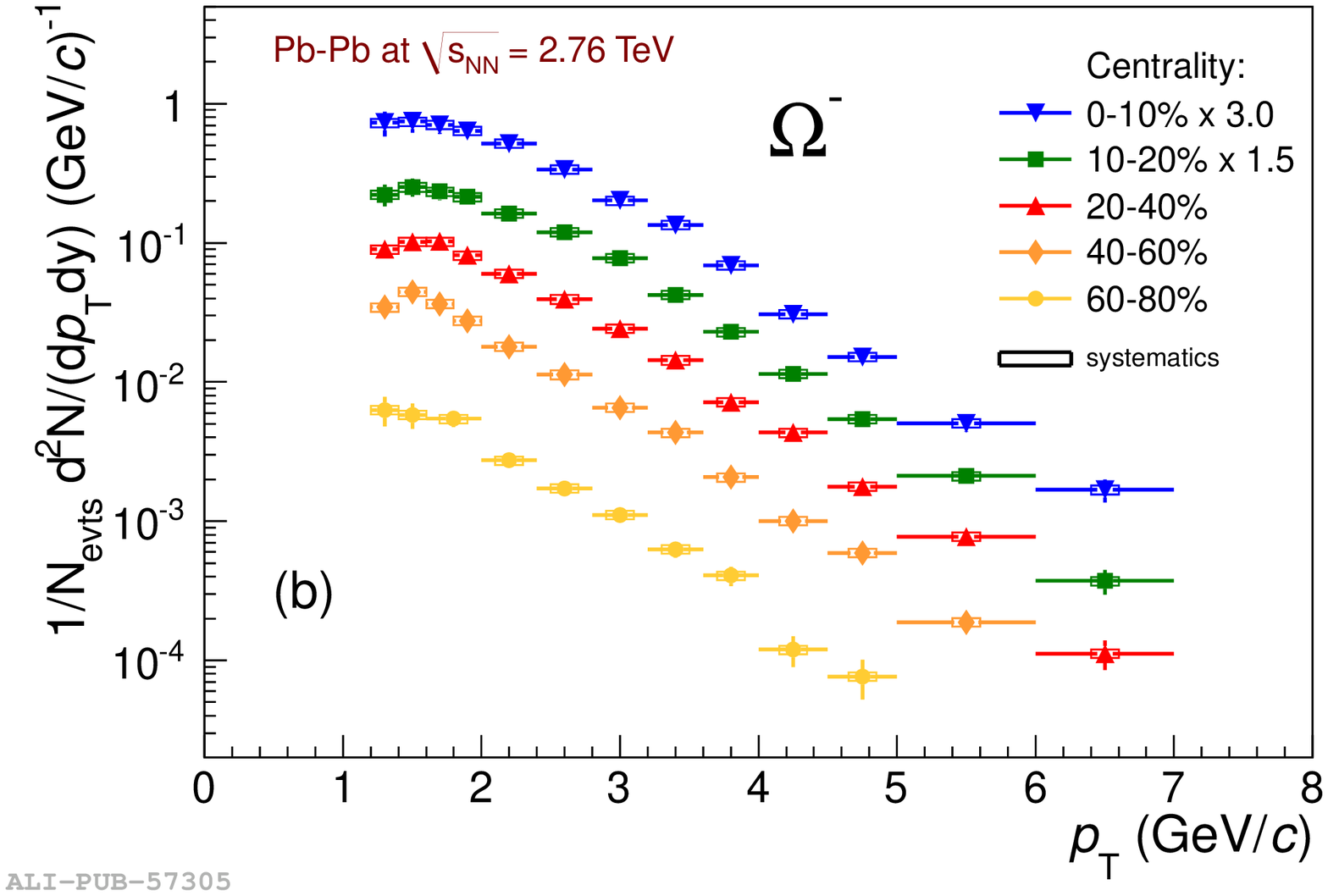, width=6.0cm}}
\end{minipage}
\vspace{0.cm}
\begin{minipage}[b]{0.49\linewidth}
\centerline{\epsfig{file=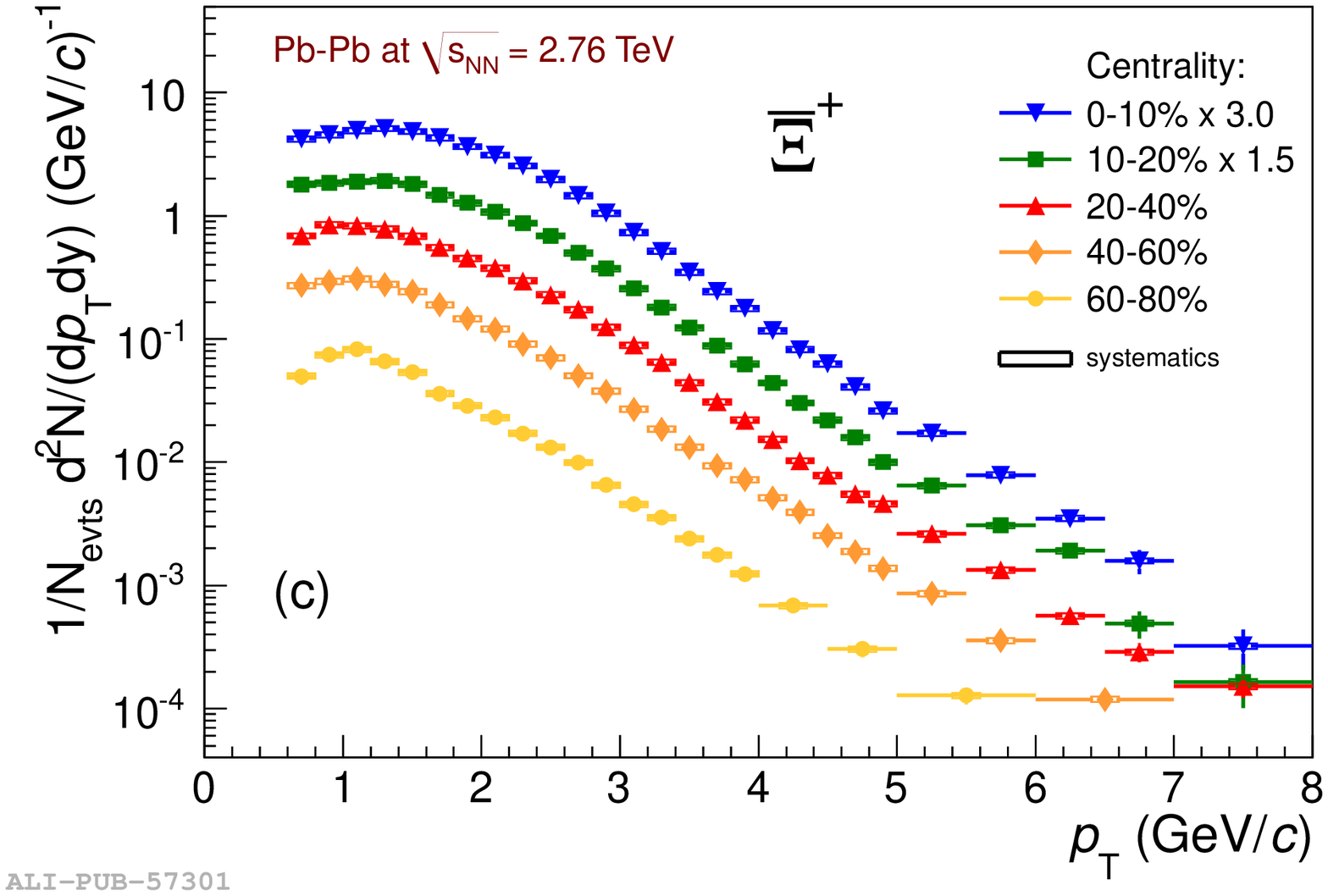, width=6.cm}}
\end{minipage}
\hspace{0.cm}
\begin{minipage}[b]{0.49\linewidth}
\centerline{\epsfig{file=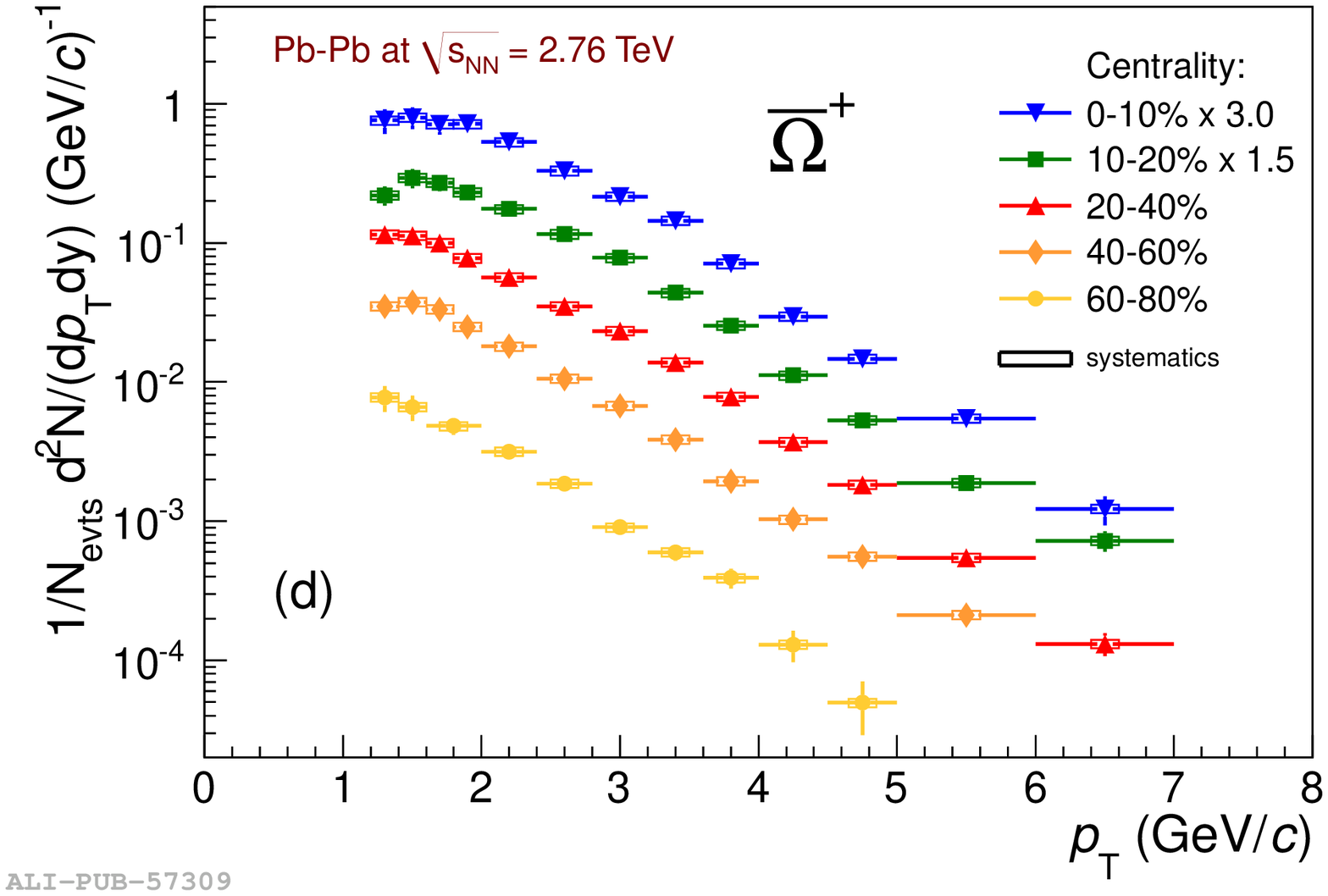, width=6.0cm}}
\end{minipage}
\vspace*{8pt}
\caption{Transverse momentum spectra for $\Xi^{-}$ and $\Omega^{-}$ (a, b)
and their anti-particles (c, d) at mid-rapidity in different centrality 
intervals for Pb-Pb collisions at $\sqrt{s_{\rm NN}} = 2.76$ TeV.
\label{fig2}}
\end{figure}

The signal was extracted, for each particle in each $p_{\rm T}$ interval,
by subtracting from the peak population the background, 
which was estimated by a
first or second order polynomial fit. Acceptance and efficiency
was estimated via a Monte-Carlo study and applied
to correct the signals. The $\Lambda$ yield was further corrected for
feed-down contributions from weak decays of $\Xi^{-}$  and $\Xi^{0}$ 
($\Omega$ contribution being negligible).  
The $p_{\rm T}$ spectra of strange and multi-strange particles
are shown for different centrality classes in Pb-Pb collisions
at $\sqrt{s_{\rm {NN}}} = 2.76$ TeV in Fig.~\ref{fig1} and~\ref{fig2},
respectively. 
%Also shown in the right panels of Fig.~\ref{fig1}
%are the resulting curves of the blast-wave fit\cite{schn93}.

\section{$\Lambda/\rm{K}^0_{\rm S}$ Ratio}

When compared with the peripheral and pp results, 
$\Lambda/\rm{K}^0_{\rm S}$ and p$/\pi$ ratios were observed to be enhanced  
at intermediate $p_{\rm T}$ in central heavy-ion collisions 
at both RHIC\cite{adle03,adam06,arse07} and SPS\cite{schu06}. 
These observations suggested that other hadronization mechanisms, 
such as coalescence, 
may open up in the deconfined medium created in heavy-ion collisions. Indeed, 
the so-called ``baryon anomaly'' could be qualitatively explained if hadrons were formed 
by recombination of two or three soft quarks\cite{frie04,hwa04}. 

\begin{figure}[!htb]
\begin{minipage}[b]{0.49\linewidth}
\centerline{\epsfig{file=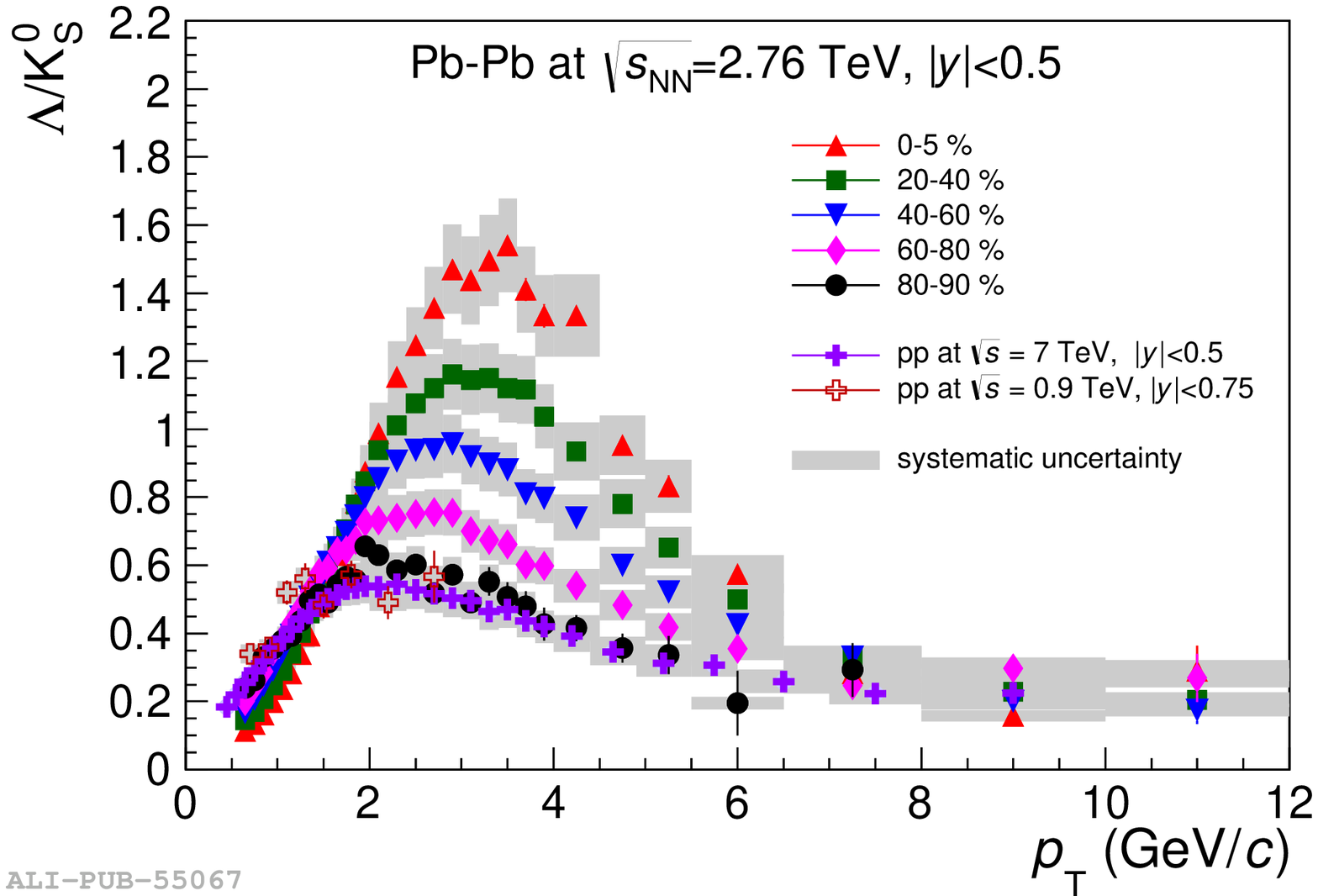, width=6.5cm}}
\end{minipage}
\hspace{0.cm}
\begin{minipage}[b]{0.49\linewidth}
\centerline{\epsfig{file=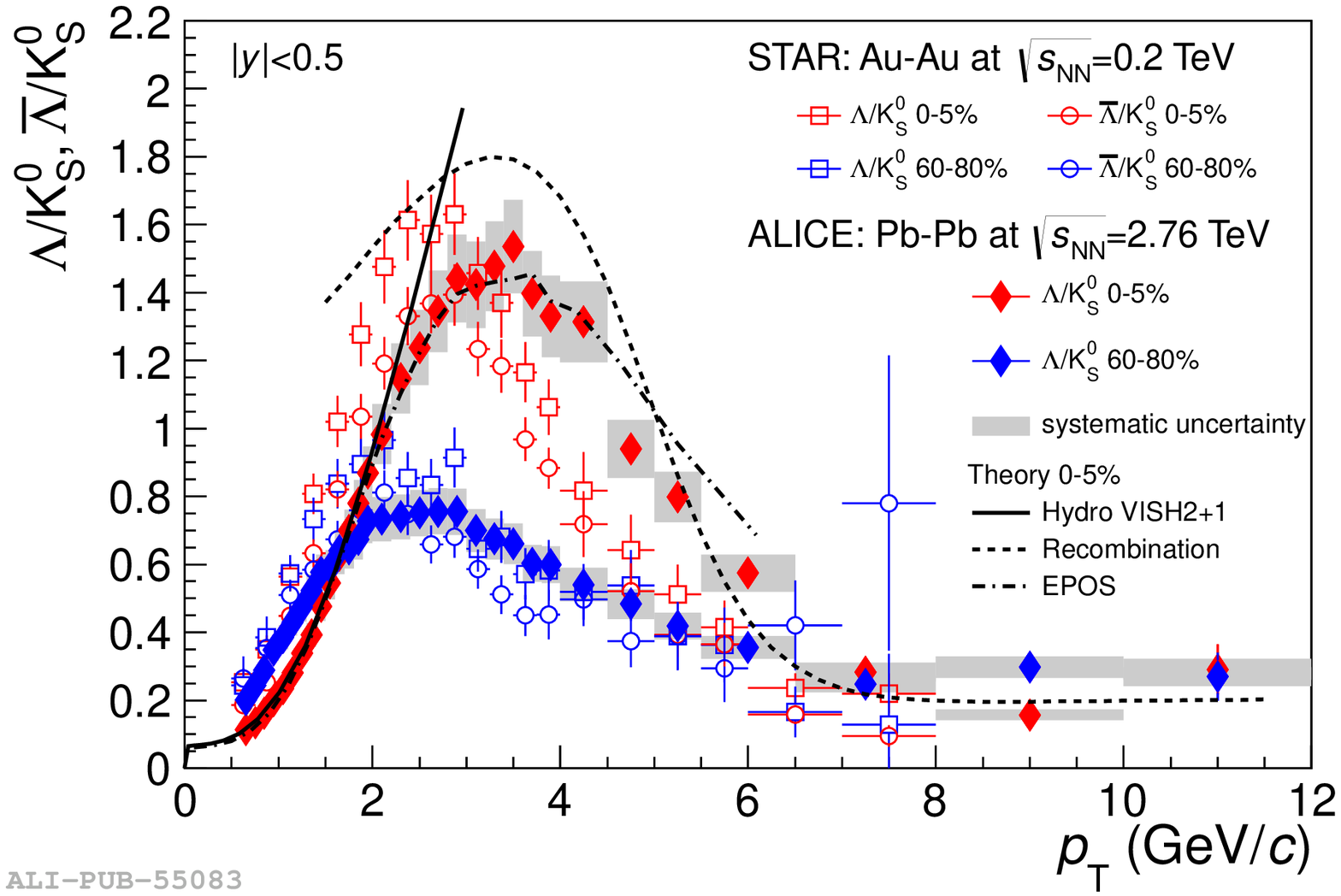,width=6.5cm}}
\end{minipage}
\vspace*{8pt}
\caption{Left: $\Lambda/\rm{K}^0_{\rm S}$ ratios for different centrality 
classes in Pb-Pb collisions at $\sqrt{s_{\rm {NN}}} = 2.76$ TeV 
are compared to the measurements in pp collisions at $\sqrt{s} = 0.9$ 
and 7 TeV. Right: $\Lambda/\rm{K}^0_{\rm S}$ ratios for central and peripheral 
Pb-Pb collisions are compared to theoretical predictions 
and measurements at RHIC energy.
\label{fig3}}
\end{figure}

Figure~\ref{fig3} (left) shows the $\Lambda/\rm{K}^0_{\rm S}$ ratios as a 
function of $p_{\rm T}$ for different centralities in Pb-Pb collisions 
at $\sqrt{s_{\rm{NN}}} = 2.76$ TeV\cite{abel13}. For comparison, also shown in the plot 
are the ratios measured in pp collisions at $\sqrt{s} = 0.9$ and 7 TeV. 
The ratio in pp collisions always stays far below 1 and appears to be 
insensitive to the 
collision energy. While the ratio in the most peripheral Pb-Pb collisions 
is compatible with that measured in pp collisions, at intermediate $p_{\rm T}$ 
it increases with collision centrality and develops a maximum at $p_{\rm T} \sim 3$ GeV/$c$ 
reaching a value of about 1.6 for the 0-5\% most central Pb-Pb collisions. 
However, at $p_{\rm T} > 7$ GeV/$c$, the $\Lambda/\rm{K}^0_{\rm S}$ ratio appears to be independent of 
collision centrality and very similar to that measured in pp collisions. This suggests that 
the production of $\Lambda$ and $\rm{K}^0_{\rm S}$ at high $p_{\rm T}$, 
even in central Pb-Pb collisions, could be dominated by vacuum-like fragmentation of energetic partons.

A comparison between LHC and RHIC measurements on $\Lambda/\rm{K}^0_{\rm S}$ ratios
is shown in the right panel of Fig.\ref{fig3} for the most central (0-5\%) and peripheral (60-80\%)
nucleus-nucleus collisions. Both $\Lambda/\rm{K}^0_{\rm S}$ and $\bar{\Lambda}/\rm{K}^0_{\rm S}$
are plotted for the STAR measurements in Au-Au collisions at $\sqrt{s_{\rm {NN}}} = 200$ GeV
considering that the anti-baryon/baryon ratio is about 0.8 instead of $\sim 1$ observed at LHC.
As can be seen from the comparison, although the magnitude of the ratio is similar, the position
of the maximum seems to shift towards higher $p_{\rm T}$ as the collision energy increases, suggesting 
a stronger radial-flow effect at higher collision energy.

Also shown in the right panel of Fig.~\ref{fig3} are theoretical model calculations 
for the most central 
collisions. As can be seen, the viscous hydrodynamical model\cite{song08} describes 
the data well up to 
$p_{\rm T}$ about 2 GeV/$c$ but deviates progressively at higher $p_{\rm T}$. 
A recombination model calculation\cite{frie08} can approximately reproduce the shape, 
but overestimates the magnitude of the data by about 15\%. The EPOS model\cite{wern12}, 
which takes into account the interaction between jets 
and the hydrodynamical medium, 
appears to describe the data reasonably well. 

\section{Strangeness Enhancement}

Strangeness enhancement, defined as the enhancement of 
the relative yield per participant 
in nucleus-nucleus collisions to that in pp 
or p-Be collisions, was proposed in the 1980s as a signature of a phase transition to 
quark-gluon plasma which was expected to take place in relativistic
nucleus-nucleus collisions\cite{rafe82}. Indeed, the enhancement of strangeness was observed 
in heavy-ion collisions at the SPS\cite{ande99,anti06,anti10} and RHIC\cite{abel08}. 
In particular, the enhancement is more pronounced for multi-strange baryons and 
decreases as the collision energy increases. 

\begin{figure}[!hbt]
\centerline{\epsfig{file=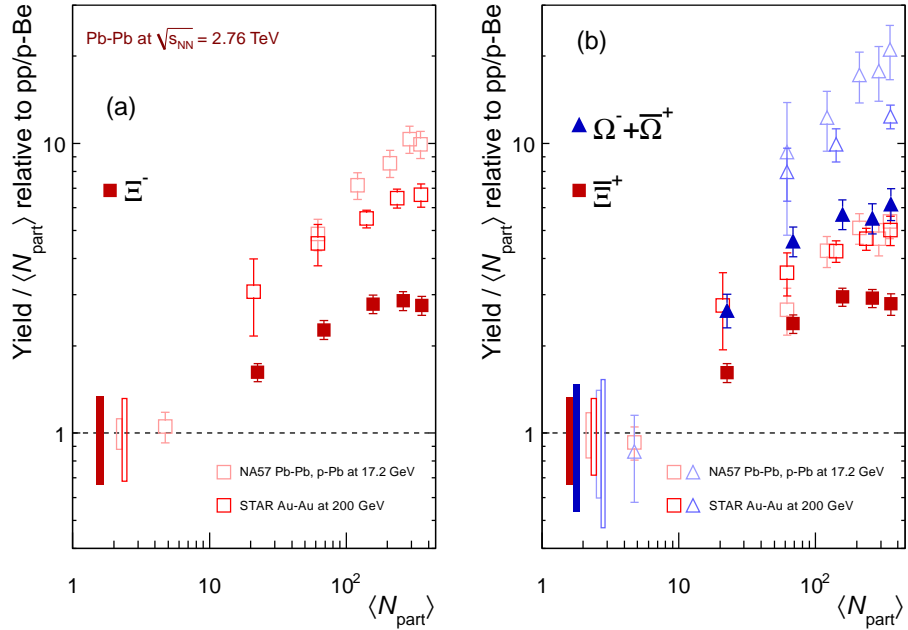, width=13.cm}}
\vspace*{8pt}
\caption{Strangeness enhancements at mid-rapidity ($|y| < 0.5$) as a function 
of the mean number of participants $\left<N_{\rm {part}}\right>$. Boxes on the dashed 
line at unity indicate statistical and systematic uncertainties 
on the pp and p-Be reference.
\label{fig4}}
\end{figure}

Figure~\ref{fig4} shows the enhancements for multi-strange baryons in Pb-Pb collisions at 
$\sqrt{s_{\rm{NN}}} = 2.76$ TeV, as a function of the mean number of participants, 
in comparison with the corresponding measurements by NA57 at SPS\cite{anti10} and 
STAR at RHIC\cite{abel08}. The enhancement increases with centrality 
and with the strangeness content of the particles as observed already at lower energies, 
but decreases with increasing energy, following the trend observed at lower energies. 
It is worth mentioning that the production of multi-strange particles 
in heavy-ion collisions does increase with collision energy 
from the SPS to the LHC. However, the increase of the multi-strange yields 
in smaller colliding systems (pBe or pp) used as reference, appears to be slightly 
faster\cite{abel12}, resulting in the less pronounced enhancement than at lower energies.

\section{Nuclear Modification Factor}

When an energetic parton traverses the hot dense QCD medium created in relativistic 
heavy-ion collisions, it suffers large energy loss due to gluon radiation and multiple 
scatterings. This parton energy loss is expected to lead to a modification of 
energetic jets (jet quenching)\cite{wang92}, which should be reflected in the 
$p_{\rm T}$ spectra of hadrons, originating from the energetic 
partons produced in initial hard collisions. To quantify the medium modification 
of the measured hadron yield in nucleus-nucleus (A-A) collisions, it is compared
to the expectation from an independent superposition of nucleon-nucleon collisions
(binary collision scaling) by introducing the nuclear modification factor:
\begin{equation}
R_{\rm{AA}}(p_{\rm{T}})=
\frac{d^2N^{\rm{AA}}/dydp_{\rm{T}}}{<T_{\rm{AA}}>d^2\sigma^{\rm{NN}}/dydp_{\rm{T}}},
\end{equation}
where $N^{\rm{AA}}$ is the particle yield in A-A collisions, 
$d^2\sigma^{\rm{NN}}/dydp_{\rm{T}}$ the cross-section of particle production in pp collisions,
and $T_{\rm{AA}}$ the geometric nuclear overlap function. 
In the absence of any nuclear modification to the incoherent hard processes, 
the nuclear modification factor at high $p_{\rm T}$ is expected to be unity 
according to the binary collision scaling.

\begin{figure}[!hbt]
\begin{minipage}[b]{0.49\linewidth}
\centerline{\epsfig{file=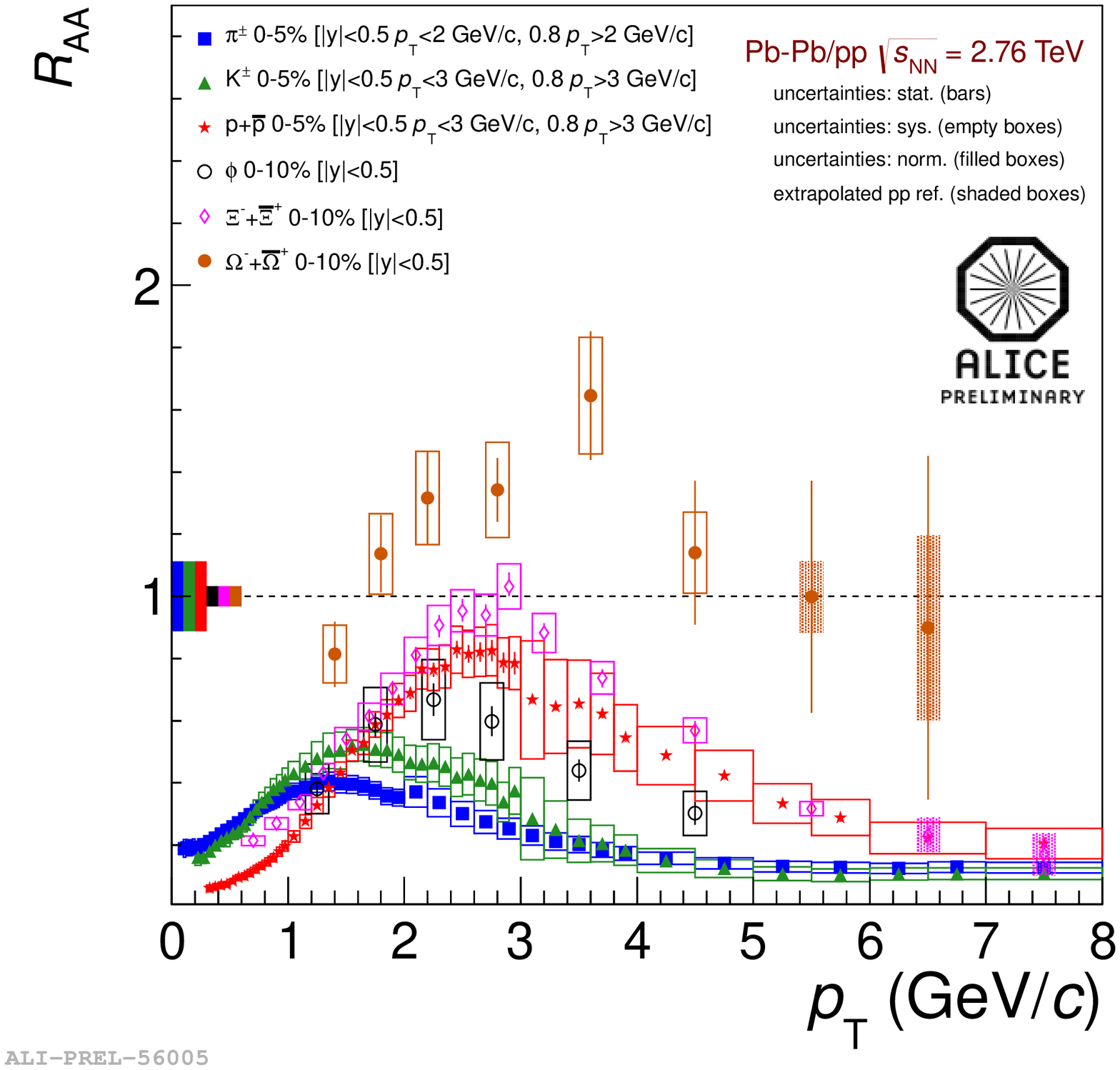, width=6.cm}}
\end{minipage}
\hspace{0.cm}
\begin{minipage}[b]{0.49\linewidth}
\centerline{\epsfig{file=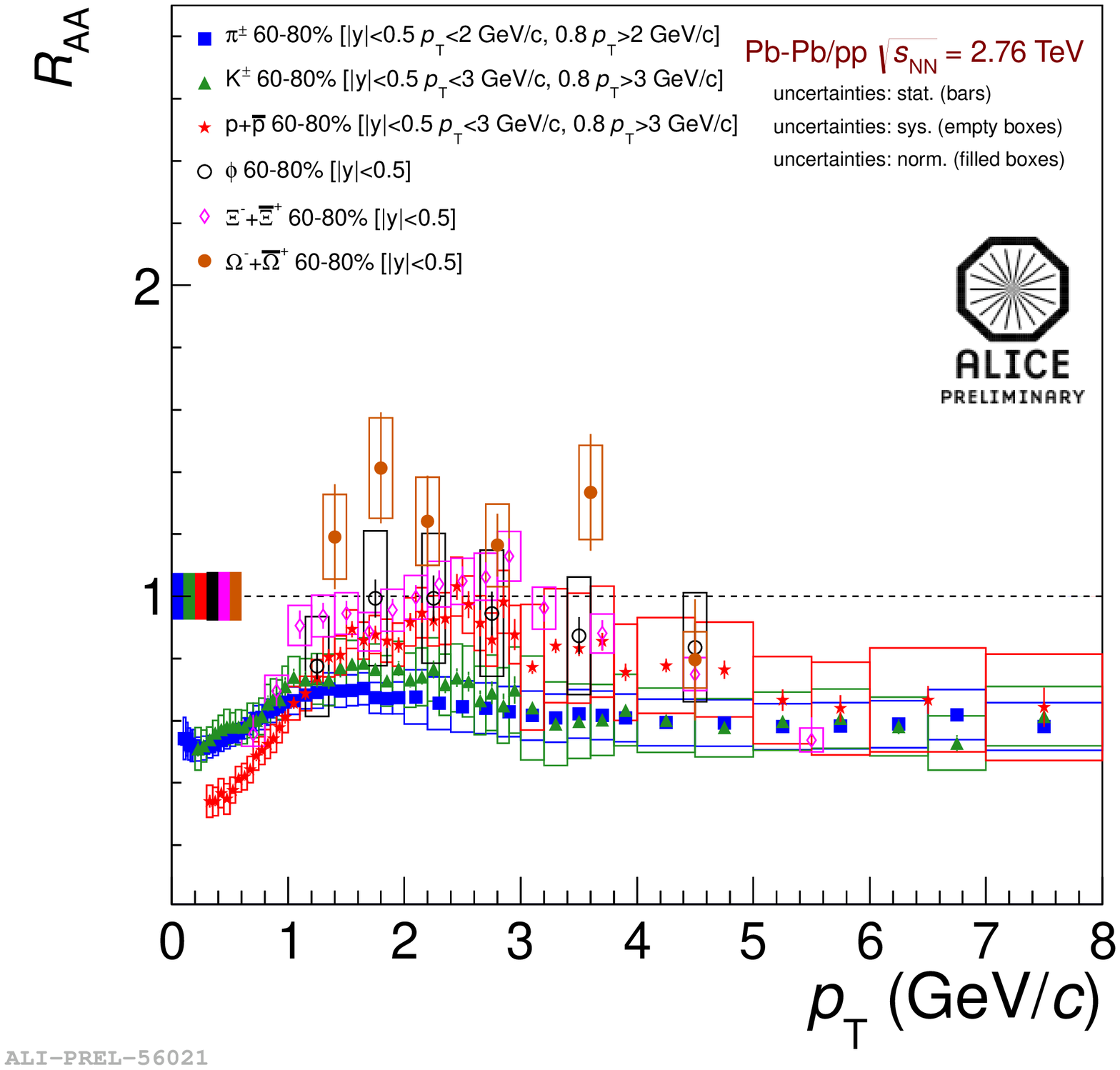, width=6.0cm}}
\end{minipage}
\vspace*{8pt}
\caption{Nuclear modification factor for $\Xi$ and $\Omega$ compared with $\pi$, K, p and $\phi$ 
in most central (left) and the most peripheral Pb-Pb collisions 
at $\sqrt{s_{\rm{NN}}} = 2.76$ TeV.
\label{fig5}}
\end{figure}

Figure~\ref{fig5} shows $R_{\rm{AA}}$ as a function of $p_{\rm T}$ for 
multi-strange baryons in most central and most peripheral Pb-Pb collisions at 
$\sqrt{s_{\rm{NN}}} = 2.76$ TeV. For comparison, also plotted in the figures 
are the nuclear modification factors for charged pions, kaons, protons and $\phi$-mesons.
At high $p_{\rm T}$ the $R_{\rm{AA}}$ for $\Xi$ seems to follow 
the same trend of the proton, whereas the $R_{\rm{AA}}$ for $\Omega$ 
appears not to be suppressed. 
At intermediate $p_{\rm T}$ a mass-ordering is clearly observed 
among the baryons and mesons, respectively.

\section{Summary}

The results of several studies on strangeness production are presented to investigate 
the properties of the strongly interacting matter created at the LHC in 
Pb-Pb collisions at $\sqrt{s_{\rm{NN}}} = 2.76$ TeV. At intermediate $p_{\rm T}$ 
$\Lambda$ production relative to $\rm{K}^0_{\rm{S}}$ is strongly enhanced in central 
Pb-Pb collisions. At high $p_{\rm T}$ however, $\Lambda/\rm{K}^0_{\rm S}$ appears to be 
similar to pp results, indicating that vacuum-like fragmentation dominates there. 
The enhancements of multi-strange baryons relative to pp increase both with the centrality 
and with the strangeness content of the baryon, but decrease with increasing energy, 
confirming the trend observed at lower energies. The comparison of 
the nuclear modification factors for multi-strange baryons to lighter particles shows that, 
while the $R_{\rm{AA}}$ for $\Xi$ at high $p_{\rm T}$ seems to follow
the same trend as of the proton, the behavior of $\Omega$ is very different 
from the others.

\section*{Acknowledgments}

This work is supported partly by the NSFC (10975061, 1137071 and 11020101060),
the National Basic Research Program of China (2013CB837803), 
and the Basic Research Program of CCNU (CCNU13F026).

%\begin{thebibliography}{000} %for 3 digits
%\begin{thebibliography}{00}  %for 2 digits

\end{document}